**Title: Uncovering networks amongst stocks returns by studying nonlinear interactions in high frequency data of the Indian Stock Market using mutual information.**

Charu Sharma[1]* and Amber Habib[1]

[1]*Department of Mathematics, Shiv Nadar University, Gautam Buddha Nagar, Uttar Pradesh, India*

*Email Id\*: charu.sharma@snu.edu.*

*Phone\*: +91-9911750311*

# Title: Uncovering networks amongst stocks returns by studying nonlinear interactions in high frequency data of the Indian Stock Market using mutual information.

In this paper, we explore the detection of clusters of stocks that are in synergy in the Indian Stock Market and understand their behaviour in different circumstances. We have based our study on high frequency data for the year 2014. This was a year when general elections were held in India, keeping this in mind our data set was divided into 3 subsets, pre-election period: Jan-Feb 2014; election period: Mar-May 2014 and :post-election period: Jun-Dec 2014. On analysing the spectrum of the correlation matrix, quite a few deviations were observed from RMT indicating a correlation across all the stocks. We then used mutual information to capture the non-linearity of the data and compared our results with widely used correlation technique using minimum spanning tree method. With a larger value of power law exponent $\alpha$, corresponding to distribution of degrees in a network, the nonlinear method of mutual information succeeds in establishing effective network in comparison to the correlation method. Of the two prominent clusters detected by our analysis, one corresponds to the financial sector and another to the energy sector. The financial sector emerged as an isolated, standalone cluster, which remain unaffected even during the election periods.

## Introduction

High frequency trading is buying and selling of large number of stocks in a very short interval of time, within fractions of seconds. With the advancement in computing and technology, nowadays it possible for the investors to carry out such trades using algorithmic trading. Based on their own strategies, investors write computer programmes, which identify trading opportunities and execute the trade with minimal human intervention. A good trading strategy should be equipped to understand the movement of stocks even at tick-by-tick level. Among various factors, which influence the change in the stock prices, change in the prices of other stocks is one of the most significant. Over the years, many researchers such as Laloux(1999), Pan & Sinha(2007), have used RMT on the empirical correlation matrix to understand the co-movements of the stocks based on daily rates of return. Spectrum of the empirical correlation matrix is studied and any deviations from the Marchenko-Pastur distribution is used to study interactions among the stocks. Pletou(2001) studied the cross correlation matrix of stock returns in a developed country and later Pan & Sinha(2007) studied the cross correlation matrix of stock returns both in developed and developing country, namely USA and India. Pan & Sinha(2007) found that an emerging market like India shows stronger interactions in price movements as compared to developed market like USA. Their entire analysis however was based on the daily rate of returns of the stocks. In this paper, we study the spectrum of the correlation matrix but now at high frequency level, 30 second time interval, with respect to Indian market. Most of the studies based on RMT performed on different exchanges suggest that bulk of the eigenvalues are in agreement with the Marchenko-Pastur distributions with few exceptions. The large eigenvalues which deviate from the Marchenko-Pastur distributions are studied to understand the influence of the market as a whole and also sectorial effect. However, correlation coefficient is a measure of linear relation between the variables. In the case of any non-linear relationship, correlation coefficient may not be able to capture this and thus most of the eigenvalues seen in agreement of the RMT maybe an artefact of this. We believe that the interactions amongst the stocks is a more complex phenomenon, and therefore simple linear techniques might not be able to capture complex nonlinearity of the data. Thus, there is a great need to develop methods, which are able to capture the non-linearity amongst the stocks at a high frequency level.

Mutual information is one such measure that quantifies the nonlinear relationship amongst the random variables that linear methods based on correlation coefficient might fail to capture. Researchers in the past have used methods based on mutual information in building biological networks Song, Langfelder, Horvath(2012), Wang, Huang(2014). Very recently researchers have started studying the stock networks based on mutual information along with their topological properties, Fiedor P.(2014); Tao, Fiedor, Holda(2015); Guo, Zhang, Tian(2018).

Estimating mutual information with a good accuracy is an important research field in itself Kraskov, Stogbauer, Grassberger(2004); Cellucci, Albano, Rapp(2005). In the past various numerical algorithms have been proposed to estimate mutual information accurately and efficiently. Cellucci, Albano, Rapp compared some of these algorithms in terms of efficiency and accuracy. Their analysis showed that Fraser-Swinney algorithm is the best in terms of accuracy but takes quite a long time. Adaptive partition method takes about 0.5% of the calculation time required by Fraser-Swinney and its accuracy is also better in comparison to uniform bin method. We used adaptive partition method to estimate mutual information on 30 second data. The remaining part of the paper is as follows. Section 2 gives the description of the data used in our analysis. Section 3 gives an overview of the methods and the methodology in our analysis. In section 4, we give the comparative study of the linear and non-linear methods using topological features of the networks. In section 5 we conclude by highlighting the salient observations and interpretations as a result of our analysis.

**Data Description**

We obtained tick-by-tick data for the year 2014 from the National Stock Exchange, India. The data was filtered to get all the stocks listed in CNX100 during that year.11 stocks were dropped from the analysis due to insufficient data values or missing data. The CNX100 index consists of the Nifty50 and the CNX Nifty Junior stocks. Table 1 and Figure 1 gives the details of the composition. The market opens at 9 o'clock in the morning and is functional till 4 PM. The trades start picking up in the first half an hour, while the last half an hour shows some ambiguity or incompleteness in the data. Considering this, we have used the data between 9:30AM and 3:30PM in our analysis. Every 30-second interval is considered a tick, and thus in each day we have 720 tick points for each stock. For the $kth$ stock, we calculate the volume weighted average price (VWAP), $\widehat{S_t^k}$, per 30 seconds and use it to find log returns per 30 seconds,

$$\widehat{S_t^k} = \frac{\sum_i v_i^k S_i^k}{\sum_i v_i^k} \tag{1}$$

Here, $v_i^k$ is the volume of the kth stock at the actual tick $i$ and $S_i^k$ is the stock price at the tick $i$ in 30 second widow at time t. The log return at time t is then calculated using equation 2.

$$R_t^k = \ln\left(\widehat{S_{t+1}^k}\right) - \ln(\widehat{S_t^k}) \tag{2}$$

Also, we specifically took year 2014 for our analysis as this was the year when general elections took place in India and a change in government was seen. We wanted to analyse the effect of this major event on the network. For this purpose, we divided our data into three parts: (a) pre-election period: Jan-Feb 2014 (b) election period: Mar-May 2014, (c) post-election period Jun-Dec 2014. Since promotional rallies took place in the month of March, elections in the month of Apr and declaration of results in the month of May, so we considered Mar-May as the election period. Table 2 summarizes the details of the data.

**Methods and Methodology**

*RMT approach on correlation coefficient matrix*

Correlation coefficient between two random variables measures the strength of linear relationship between them.

$$\text{Let } R = [R^1, R^2, \ldots, R^{89}] \tag{3}$$

$$\text{where } R^k = \begin{bmatrix} R_1^k \\ R_2^k \\ R_3^k \\ . \\ . \\ . \\ R_m^k \end{bmatrix}, k = 1,2,\ldots,89 \tag{4}$$

$R_t^k$ is given by equation (2).

We then calculate the correlation between each pair of stocks and build the correlation matrix $C$ of size $89 \times 89$ where the $(i,j)th$ entry in the matrix denotes the correlation between the $ith$ stock and $jth$ stock. We do it separately for all 3 time spans, pre-election, election and post-election. In Figure 2 we plot the boxplot corresponding to the distribution of correlation coefficients for these 3 different time span. During the election period more pairs are seen to have higher correlation coefficient in comparison to pre-election and post-election periods. This phenomenon can be easily understood as a fact that there is a lot of news, speculations moving around in the market which in a way influence the stock market. Thus observing high correlation coefficient during election time could be due to market effect, which was later established when we analysed $C$ under RMT.

The statistical properties of correlation matrix are well established in literature []. With large number of variables and large number of sample points, and under the hypothesis that $C$ is a random correlation matrix, the distribution of eigenvalues of $C$ can very well be approximated by Marchenko-Pastur distribution. For large $k$ and $m$, i.e. $k \to \infty, m \to \infty$ such that $Q = \frac{m}{k}$ is fixed, the probability distribution of the eigenvalues of a random correlation matrix is given by

$$f_{rm}(\lambda) = \frac{Q}{2\pi} \frac{\sqrt{(\lambda_{max}-\lambda)(\lambda-\lambda_{min})}}{\lambda} \tag{5}$$

where $\lambda_{max}$ and $\lambda_{min}$ are the maximum and minimum eigenvalue of $C$ given by

$$\lambda_{max} = \left(1 + \sqrt{\frac{1}{Q}}\right)^2 \text{ and } \lambda_{min} = \left(1 - \sqrt{\frac{1}{Q}}\right)^2 \tag{6}$$

Any deviations from this distribution points towards the rejection of the null hypothesis that entries in $C$ are random. Figure 3 gives comparison of empirical pdfs and theoretical pdfs of the eigenvalues of $C$. Table 3 summarizes the statistics from the empirical and theoretical distribution. More than 40% of the data was seen to be deviated from the Marchenko-Pastur distribution in all three-time span. Since eigenvectors corresponding to large eigenvalues, also known as principal components carry useful information in comparison to eigenvectors corresponding to small eigenvalues and thus we are interested in analysing only the large eigenvalues, which are deviating from the RMT.

Also, in order to show that the deviations from the RMT is not because of the finite number of variables, 89 in our case, we tested this procedure on the surrogate data generated by randomly shuffling the returns for each stock. Figure 4 gives the comparison of the empirical pdfs and theoretical pdfs of the eigenvalues of $C$ corresponding to the testing data. It is quite evident that testing data matches well with the Marchenko-Pastur distribution, indicating that the deviations from this distribution in the original data are genuinely due to the correlation between the stocks.

*Mutual Information Method*

Mutual Information between two random variables captures mutual dependence between them. Correlation coefficient helps to determine whether there is a linear relationship amongst the variables, on the other hand mutual information helps to measure the non-linear relationship between the variables. We strongly believe that interactions amongst the stocks are non-linear and thus understanding these interactions using mutual information is more appropriate. In information theory, Shannon Entropy is a measure of "uncertainty" or "unpredictability" of a random variable or a random vector. For discrete random variables $X$ and $Y$, their joint entropy is defined as

$$H(X,Y) = -\sum_i \sum_j f_{X,Y}(x_i, y_j) \log\left(f_{X,Y}(x_i, y_j)\right) = E[-\log(f_{X,Y})] \qquad (7)$$

where $f_{X,Y}(x_i, y_j)$ is joint probability mass function of $X$ and $Y$. Also entropy of a discrete random variable with probability mass function $f_X$ is defined as

$$H(X) = -\sum_i f_X(x_i)\log(f_X(x_i)) = E[-\log(f_X)] \qquad (8)$$

Mutual Information of discrete random variables X and Y is defined as

$$I(X,Y) = H(X) + H(Y) - H(X,Y) = \sum_i \sum_j f_{X,Y}(x_i, y_j) \log\left(\frac{f_{X,Y}(x_i, y_j)}{f_X(x_i) f_Y(y_j)}\right) \qquad (9)$$

A generalization to continuous case is

$$I(X,Y) = \iint f_{X,Y}(x, y) \log\left(\frac{f_{X,Y}(x,y)}{f_X(x) f_Y(y)}\right) dx dy \qquad (10)$$

Cellucci, Albano, Rapp(2005) gave comparative study of methods to estimate mutual information in case of continuous random variables. We used non-uniform adaptive partition algorithm over Fraser-Swinney algorithm to estimate mutual information as it was seen best in terms of both efficiency and accuracy. Based on mutual information, the distance between two stocks $k$ and $s$ is defined as

$$d(R^k, R^s) = 1 - \frac{I(R^k, R^s)}{H(R^k, R^s)} \qquad (11)$$

We can check it satisfies all properties of a metric.

*Minimum Spanning tree*

For connected graphs, a spanning tree is a subgraph that connects every node in the graph with no cycles. There may exist more than one spanning tree for any given graph. If the weights are assigned to each edge, then minimum spanning tree is the spanning tree whose edges have the least total weight. To build a MST stock network, we quantify the distance between each pair of stocks and use this distance as the edge weight between each pair stocks. In our analysis we have considered two models on stocks, one based on linear relationship between stocks using correlation coefficient and other based on on-linear relationship using mutual information. For both the cases, we defined measure of distance between pairs of stocks, equation 11 and 12, and used them to construct MSTs. There are two well known models to construct MST, Kruskal algorithm and Prim algorithm. We used Prim algorithm in our computations.

**Analysing the networks**

*Comparative study of the methods*

In the past, many researchers have used correlation coefficient of daily rate of return of stocks to understand the networks amongst them. (Mantegna, 1999). Plerou (2001) and Pan & Sinha (2007). Pan and Sinha studied daily rate of return in context of Indian stock market and they observed strong correlation movement in comparison to developed country like US. In their study, they observed bulk of the data in synergy with RMT with few deviations that indicate market effect. In our analysis, we worked on high frequency data at a scale of 30 second. Around 42%, 50% and 58% deviations were observed from the RMT during pre-election, election and post-election period respectively out of which around 7%, 9% and 8% deviations corresponds to large eigenvalues respectively. Figure 5 gives the eigenvectors corresponding to three largest eigenvalues (also known as principal components) corresponding to different time span. Stocks corresponding to financial sector and the IT sectors have an edge over other sectors in the first eigenvector over all three-time spans, pre-election, election and post-election. Also from each sector, there are few stocks, which are dominant contributors to the first eigenvector. During the pre-election time, financial sector, IT sector and energy sector becomes key contributors in second eigenvector. However, during the election and post- election it is the financial sector and the energy sector, which are the dominant contributors towards second eigenvector.

Deviations from the RMT suggests that the correlations observed are not all due to randomness and hence in order to study the linear relationship between the stocks we constructed minimum spanning tree graph with respect to the distance metric as,

$$d(R^k, R^s) = \sqrt{2(1 - \rho_{R^k, R^s})}, \qquad (12)$$

where $\rho_{R^k, R^s}$ is correlation coefficient between rate of returns of the two stocks $R^k, R^s$. Figure 6, 7 and 8 gives networks based on correlation method for pre-election, election and post- election periods. To capture the non-linearity in the data, we constructed mutual information based MST using distance given in equation 9. Figure 9, 10 and 11 gives networks based on mutual information for pre-election, election and post- election periods. We used *Gephi 0.9.2* to plot these networks. In case of mutual information method, we carried out hypothesis testing at 5% level of significance and took value of mutual information between a pair of stock as zero in the case when hypothesis of a zero mutual information could not be rejected.

In order to analyse how effective is non-linear method based on mutual information in comparison to linear method based on correlation coefficient, we plotted normalized mutual information values against the correlation coefficient values of all the 3916 pairs of the stocks. Normalized mutual information between two random variables $X$ and $Y$ is defined as

$$U(X,Y) = 2\frac{I(X,Y)}{E(X)+E(Y)} \tag{13}$$

where $I(X,Y)$ is mutual information between two random variables and $E[X], E[Y]$ are their respective entropy. Figure 12 gives the plots corresponding to all three time spans, pre-election, election and post-election. We observe that in all the three cases, larger values of correlation coefficient are associated with larger values of mutual information but there are substantial number of instances when smaller values of correlation coefficient is associated with large values of mutual information. This suggest that the non-linear method based on mutual information non only managed to capture strong linear relationships but at the same time captured the non-linearity found in the data which linear method based on correlation coefficient failed to capture. Also, there are instances when magnitude of the values of mutual information are much smaller in comparison to the values of correlation (correlation coefficient$<$ 0.1). However, it was found out that such instances were fewer in all the time-spans in comparison to the instances when large values of mutual information are associated with small values of correlation. We believe this could be due to some randomness, which even mutual information method also failed to capture. In all, it is quite evident that mutual information method is much efficient in comparison to widely used correlation method to build stock network.

In order to get more insightful information, we studied some of the centrality measures like degree, degree distributions and eigenvalue centrality with respect to stock networks obtained using mutual information and correlation coefficient methods.

*Degree Distribution of the networks*

In a graph, degree of a node is the number of links attached to that node. Nodes with high degrees are important nodes in a graph and they are called hubs. Stocks corresponding to hubs in a network are the stocks who respond first in case of any inflow of the information and subsequently the information is transferred to the stocks found in periphery of these hubs. We looked at degree distribution of the stocks in all the networks. Figure 13 gives the degree distribution of all the stocks under different methods and different time span. Stocks from financial sector, like ICICIbank, PNB, Reliance, Yes Bank, are observed to have degrees more than 4, are clearly the dominant stocks irrespective of the methods and the time span. Other than financial sector, there are few stocks from the IT and Energy sector which are found to have high degree .All the stocks with degree more than 4 are seen to come under large cap companies.

Emergence of hubs in a network is seen as a property of a scale free network, i.e. a network whose degree distribution follows power law distribution, with power law exponent $\alpha$, $2 < \alpha < 3$. Since there were very few hubs seen in the degree distribution corresponding to the networks(Figure 13), we were motivated to check the scale free property in the network. In this context, we analysed the probability distribution of the degrees of the stocks in the networks based on correlation coefficient method and mutual information method, for the three time periods, pre-election, election and post-election time periods (Guo, 2018). If $f_D$ is the probability density function of the degree distribution, then we estimated it such that $f_D(x) \propto x^{-\alpha}$ for some $\alpha$ called as power law exponent. We assumed min value of $x$ as 1 and used method of maximum likelihood estimator to estimate $\alpha$. Table 4 summarizes the parameter estimation for all the 12 networks. It is quite evident that estimated value of $\alpha$ i.e. $\hat{\alpha}$ is observed to be smallest during the election time irrespective of the method used to build the network. Smaller value of $\hat{\alpha}(<2)$ is an indication that the market is violating the scale free network, i.e. there are substantial number of nodes with higher degrees and thus market is closely knitted as a network during the election time. Also, across all time span, $\hat{\alpha}$ is high in case of Mutual Information in comparison to the correlation method. Thus, again nonlinear method based on mutual information is a good choice of method to study scale free networks in Indian Stock Market at a high frequency level.

*Eigenvalue centrality measure*

We also considered eigenvalue centrality measure to identify important stocks in terms of flow of information and stocks which appear in the surroundings of these stocks. For each node, we define a relative score such that for a node, its connections to a high-scoring nodes contribute more towards its score in comparison to all the low scoring nodes in its neighbourhood. High scoring nodes are gain refereed as hubs. We realized spectrum of the adjacency matrix and Laplacian matrix help us to define such a scoring system.

    Adjacency matrix of a graph is a $n \times n$ matrix, where $n$ is number of nodes in the graph with $(i,j)th$ entry in the matrix is 1, if there is an edge between node $i$ and $j$ and it is 0 if there is no edge. Adjacency matrix is a non-negative matrix and as an application of Perron-Frobenius theorem, the eigenvector corresponding to the highest eigenvalue, also known as Perron eigenvector emerges as a good choice for the scoring

system that we are looking for. Table 5, 6 and 7 gives the hubs, nodes with high scores, and their respective normalized score in Perron eigenvector.

In order to capture the stocks, present in the neighbourhood of the stocks with high scores, we considered Fidler vector, eigenvector corresponding to the second smallest eigenvalue of the Laplacian matrix. This vector is used to detect communities within the network. We studied communities corresponding to hubs since hubs are the key players in the market, information flows through them quickly in comparison to other stocks in the market.

Financial sector companies emerged as hubs in all the three time-spans. During the election time, stocks present under financial sector and the energy sector emerged as the dominant stocks in the market(Table 6). During this time, except for few stocks, the scores based on Perron vector was observed to be uniformly distributed between the stocks which points towards the uniform market effect during the election time. The effect of major event like election is evident from this analysis. Also, post-election, companies from different sectors were seen to drive the market along with the financial sector companies and energy sector. While analysing the neighbourhood of the hubs (stocks with high scores) we observed sectorial effect i.e. most of the neighbours found in the periphery of the hubs belong to the same sectors.

**Conclusion**

The aim of this paper was to study interactions between the stocks at the tick-by-tick level with respect to the Indian stock market. For this purpose we pick 30 seconds as our tick size and study behaviour of 89 stocks out of 100 stocks listed in CNX100 for the year 2014. We analysed the spectrum of the correlation matrix to study the randomness. More than 40% of the deviations were observed from the RMT indicating that the pairwise correlation coefficients are not random. We then compared the pairwise correlation coefficients with their respective mutual information. Our analysis showed that mutual information managed to capture not only the linear relationship but also the non-linear relationship perfectly. We thus propose that networks based on mutual information, in comparison to the networks based on correlation coefficient, captures the real dynamics between the stocks at a high frequency level. Networks constructed using mutual information showed a scale-free property in comparison to correlation coefficient method. Also, on the basis of our analysis we observed, that India being a developing country, its stock market is greatly influenced by the financial sector. It was also observed, that major political event like national elections, had an influence on the stock price movements. Increase in the number of pairs with higher correlation coefficients were seen during the election time.

Based on our analysis, we finally conclude that stock networks based on the mutual information method captures the dynamics of the stock market more efficiently at high frequency level. In future, we wish to explore these networks in more depth and use them for portfolio selection at a high frequency level.

Acknowledgements: NBHM Grant

Table 1: Stocks from different industry studied in our analysis

| Industry Type | No. of Stocks studied |
|---|---:|
| INDUSTRIAL MANUFACTURING | 5 |
| CEMENT & CEMENT PRODUCTS | 5 |
| SERVICES | 2 |
| AUTOMOBILE | 10 |
| CONSUMER GOODS | 14 |
| PHARMA | 10 |
| FINANCIAL SERVICES | 14 |
| ENERGY | 10 |
| TELECOM | 3 |
| METALS | 6 |
| CONSTRUCTION | 2 |
| IT | 6 |
| CHEMICALS | 1 |
| FERTILISERS & PESTICIDES | 1 |

Table 2: Features considered for three different dataset, pre-election, election, post-election.

|  | Jan-Feb | Mar-May | Jun-Dec |
|---|---|---|---|
| No. of trading days | 42 | 46 | 141 |
| No. of sample points | 89 | 89 | 89 |
| No. of features | 30198 | 33074 | 101379 |

Table 3: Summary of comparison of empirical and theoretical distribution of eigenvalues of correlation matrix

|  | pre-election | election | post-election |
|---|---|---|---|
| $\lambda_{max}$ (theoretical) | 1.11 | 1.11 | 1.06 |
| $\hat{\lambda}_{max}$ (empirical) | 7.40 | 7.68 | 8.23 |
| $\hat{\lambda}_{max}/\lambda_{max}$ | 6.66 | 6.94 | 7.77 |
| Data in agreement with RMT(%) | 58.43% | 50.56% | 42.13% |
| Data greater than lambda max(%) | 6.74% | 8.99% | 7.87% |
| Data less than lambda min(%) | 34.83% | 40.45% | 50.00% |

Table 4: Estimated power law exponent $\alpha$, for degree distribution corresponding to 12 networks.

|  | Correlation method | Mutual Information method |
|---|---|---|
| Pre-election Jan-Feb 2014 | 1.95 | 2.02 |
| Election Mar-May 2014 | 1.93 | 1.93 |
| Post-election Jun-Dec 2014 | 1.93 | 2.01 |

Table 5: High scoring stocks on the basis of scores in Perron vector, for the election period i.e. Jan-Feb2014

| Hubs on the basis of scores in Perron vector , Jan-Feb 2014 | | | | | |
|---|---|---|---|---|---|
| correlation method | | | mutual information | | |
| Name | Business Sector | normalized score in eigenvector corresponding to largest eigenvalue | Name | Business Sector | normalized score in eigenvector corresponding to largest eigenvalue |
| ICICIBANK | FINANCIAL SERVICES | 10.85% | CONCOR | SERVICES | 14.19% |
| YESBANK | FINANCIAL SERVICES | 6.01% | BEL | INDUSTRIAL MANUFACTURING | 2.80% |
| PNB | FINANCIAL SERVICES | 2.81% | HINDPETRO | ENERGY | 2.62% |
| INDUSINDBK | FINANCIAL SERVICES | 2.50% | TATAMTRDVR | AUTOMOBILE | 2.54% |
| LT | CONSTRUCTION | 2.49% | BAJFINANCE | FINANCIAL SERVICES | 2.54% |
| RELIANCE | ENERGY | 2.49% | ABB | INDUSTRIAL MANUFACTURING | 2.46% |
| | | | ACC | CEMENT & CEMENT PRODUCTS | 2.46% |
| | | | ADANIPORTS | SERVICES | 2.46% |
| | | | BAJAJFINSV | AUTOMOBILE | 2.46% |
| | | | INFRATEL | FINANCIAL SERVICES | 2.46% |
| | | | BOSCHLTD | INDUSTRIAL MANUFACTURING | 2.46% |
| | | | BRITANNIA | CONSUMER GOODS | 2.46% |
| | | | CADILAHC | PHARMA | 2.46% |
| | | | CUMMINSIND | INDUSTRIAL MANUFACTURING | 2.46% |
| | | | DIVISLAB | INDUSTRIAL MANUFACTURING | 2.46% |
| | | | EICHERMOT | AUTOMOBILE | 2.46% |
| | | | EMAMILTD | CONSUMER GOODS | 2.46% |
| | | | GSKCONS | CONSUMER GOODS | 2.46% |
| | | | GLAXO | ENERGY | 2.46% |

Table 6: High scoring stocks on the basis of scores in Perron vector, for the election period i.e. Mar-May 2014

| Hubs on the basis of scores in Perron vector, Mar-May 2014 | | | | | |
|---|---|---|---|---|---|
| correlation method | | | mutual information | | |
| Name | Business Sector | normalized score in eigenvector corresponding to largest eigenvalue | Name | Business Sector | normalized score in eigenvector corresponding to largest eigenvalue |
| YESBANK | FINANCIAL SERVICES | 13.59% | PNB | FINANCIAL SERVICES | 9.63% |
| PNB | FINANCIAL SERVICES | 3.13% | YESBANK | FINANCIAL SERVICES | 6.96% |
| RELIANCE | ENERGY | 3.06% | BANKBARODA | FINANCIAL SERVICES | 6.36% |
| ICICIBANK | FINANCIAL SERVICES | 2.90% | TATASTEEL | METALS | 3.45% |
| TCS | IT | 2.78% | RELIANCE | ENERGY | 3.18% |
| BHEL | INDUSTRIAL MANUFACTURING | 2.68% | BHARTIARTL | TELECOM | 2.68% |
| BHARTIARTL | TELECOM | 2.68% | BHEL | INDUSTRIAL MANUFACTURING | 2.67% |
| TATAMOTORS | AUTOMOBILE | 2.68% | LT | CONSTRUCTION | 2.67% |
| TECHM | IT | 2.68% | | | |
| ADANIPORTS | SERVICES | 2.58% | | | |
| ASHOKLEY | AUTOMOBILE | 2.58% | | | |
| AUROPHARMA | PHARMA | 2.58% | | | |
| BAJAJ-AUTO | AUTOMOBILE | 2.58% | | | |
| CIPLA | PHARMA | 2.58% | | | |
| COALINDIA | METALS | 2.58% | | | |
| COLPAL | CONSUMER GOODS | 2.58% | | | |
| DLF | CONSTRUCTION | 2.58% | | | |
| HAVELLS | CONSUMER GOODS | 2.58% | | | |
| INDUSINDBK | FINANCIAL SERVICES | 2.58% | | | |
| KOTAKBANK | FINANCIAL SERVICES | 2.58% | | | |
| LICHSGFIN | FINANCIAL SERVICES | 2.58% | | | |
| LT | CONSTRUCTION | 2.58% | | | |
| M&M | AUTOMOBILE | 2.58% | | | |
| MARUTI | AUTOMOBILE | 2.58% | | | |
| PFC | FINANCIAL SERVICES | 2.58% | | | |
| TITAN | CONSUMER GOODS | 2.58% | | | |
| MCDOWELL-N | CONSUMER GOODS | 2.58% | | | |

Table 7: High scoring stocks on the basis of scores in Perron vector, for the post-election period i.e. Jun-Dec 2014

| Hubs on the basis of scores in Perron vector, Jun-Dec 2014 | | | | | |
|---|---|---|---|---|---|
| correlation method | | | mutual information | | |
| Name | Business Sector | normalized score in eigenvector corresponding to largest eigenvalue | Name | Business Sector | normalized score in eigenvector corresponding to largest eigenvalue |
| YESBANK | FINANCIAL SERVICES | 11.78% | GLAXO | PHARMA | 15.44% |
| ICICIBANK | FINANCIAL SERVICES | 4.54% | GRASIM | CEMENT & CEMENT PRODUCTS | 3.33% |
| TATASTEEL | METALS | 3.74% | ABB | INDUSTRIAL MANUFACTURING | 3.08% |
| RELIANCE | ENERGY | 2.82% | BAJFINANCE | FINANCIAL SERVICES | 3.08% |
| ADANIPORTS | SERVICES | 2.81% | BAJAJFINSV | FINANCIAL SERVICES | 3.08% |
| IDEA | TELECOM | 2.67% | BEL | INDUSTRIAL MANUFACTURING | 3.08% |
| LT | CONSTRUCTION | 2.67% | INFRATEL | TELECOM | 3.08% |
| SIEMENS | INDUSTRIAL MANUFACTURING | 2.67% | BOSCHLTD | AUTOMOBILE | 3.08% |
| TCS | IT | 2.67% | BRITANNIA | CONSUMER GOODS | 3.08% |
| TATAMOTORS | AUTOMOBILE | 2.67% | CADILAHC | PHARMA | 3.08% |
| AMBUJACEM | CEMENT & CEMENT PRODUCTS | 2.56% | COLPAL | CONSUMER GOODS | 3.08% |
| ASHOKLEY | AUTOMOBILE | 2.55% | CONCOR | SERVICES | 3.08% |
| AUROPHARMA | PHARMA | 2.44% | CUMMINSIND | INDUSTRIAL MANUFACTURING | 3.08% |
| DABUR | CONSUMER GOODS | 2.44% | EICHERMOT | AUTOMOBILE | 3.08% |
| IBULHSGFIN | FINANCIAL SERVICES | 2.44% | EMAMILTD | CONSUMER GOODS | 3.08% |
| POWERGRID | ENERGY | 2.44% | GSKCONS | CONSUMER GOODS | 3.08% |
| TATAPOWER | ENERGY | 2.44% | GODREJCP | CONSUMER GOODS | 3.08% |
| TECHM | IT | 2.44% | IBULHSGFIN | FINANCIAL SERVICES | 3.08% |
| TITAN | CONSUMER GOODS | 2.44% | MARICO | CONSUMER GOODS | 3.08% |
| UPL | FERTILISERS & PESTICIDES | 2.44% | OIL | ENERGY | 3.08% |
| MCDOWELL-N | CONSUMER GOODS | 2.44% | OFSS | IT | 3.08% |
| WIPRO | IT | 2.44% | PIDILITIND | CHEMICALS | 3.08% |
| | | | PEL | PHARMA | 3.08% |
| | | | SHREECEM | CEMENT & CEMENT PRODUCTS | 3.08% |
| | | | TORNTPHARM | PHARMA | 3.08% |
| | | | UBL | CONSUMER GOODS | 3.08% |

Figure 1. Sectorial distribution of stocks from different industry studied in our analysis.

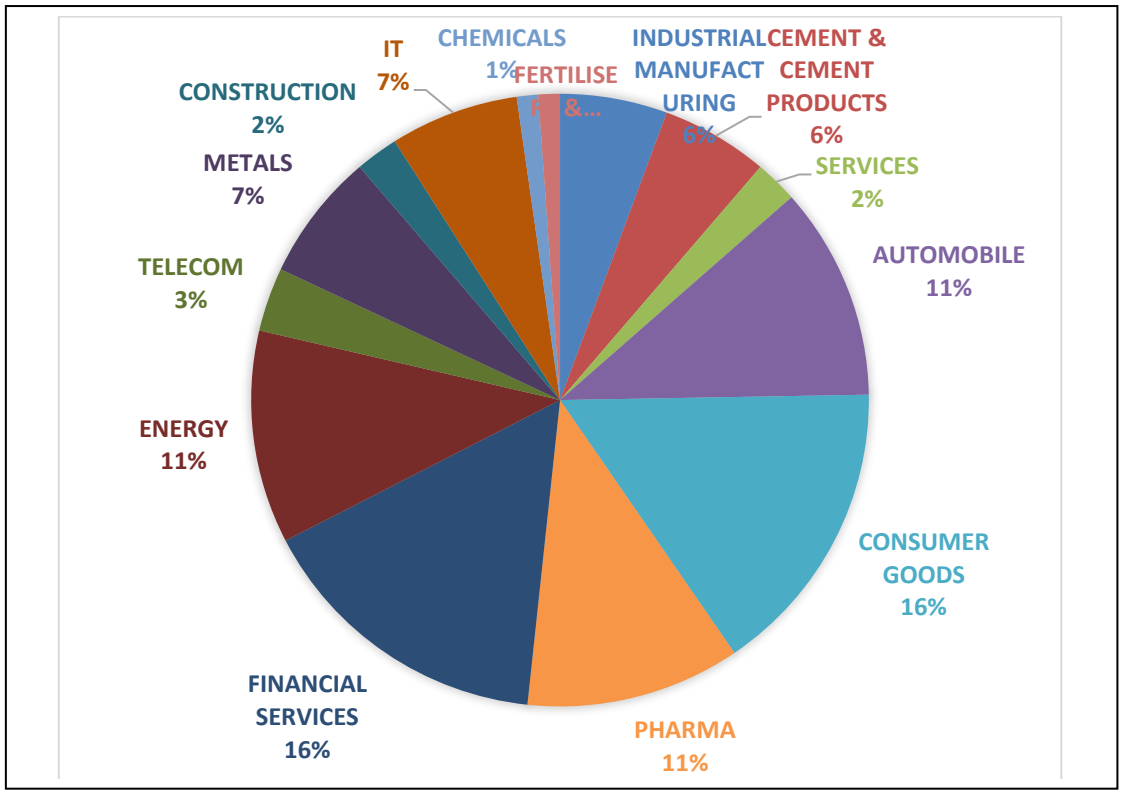

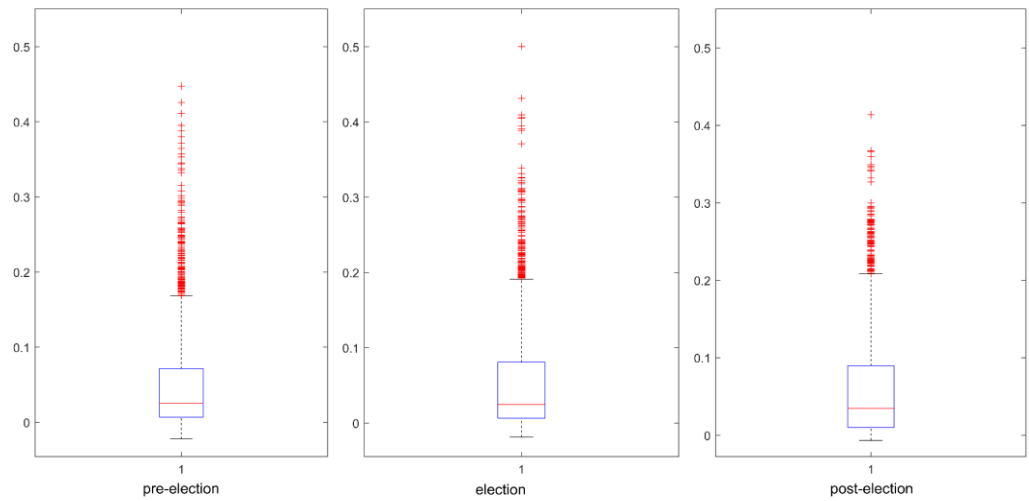

Figure 2: Distribution of correlation coefficient between all 3916 pairs of stocks during different time span, left graph corresponds to pre-election period, centre graph for election period and rightmost graph for post-election period

Figure 3: Eigenvalue distribution of correlation coefficients corresponding to (a) pre-election, (b) election and (c) post-election period. Histograms corresponds to empirical probability distribution and solid line corresponds to the theoretical pdf.

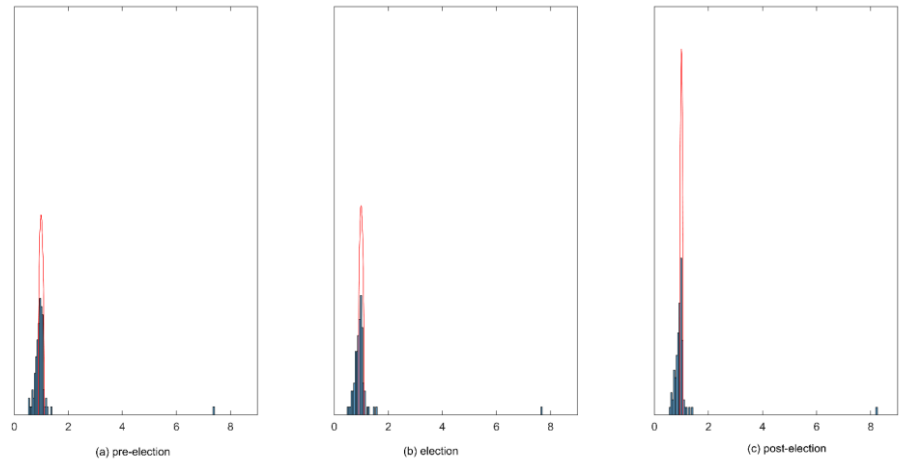

Figure 4: Eigenvalue distribution of correlation coefficients on the surrogate data obtained after reshuffling of the returns of each stock. (a), (b) and (c) are graphs corresponding to pre-election, election and post-election period by testing on 1 such surrogate dataset. (a), (b) and (c) are graphs corresponding to pre-election, election and post-election period on ensemble surrogate datasets, i.e. repeating one such trial 50 times. Histograms corresponds to empirical probability distribution and solid line corresponds to the theoretical pdf.

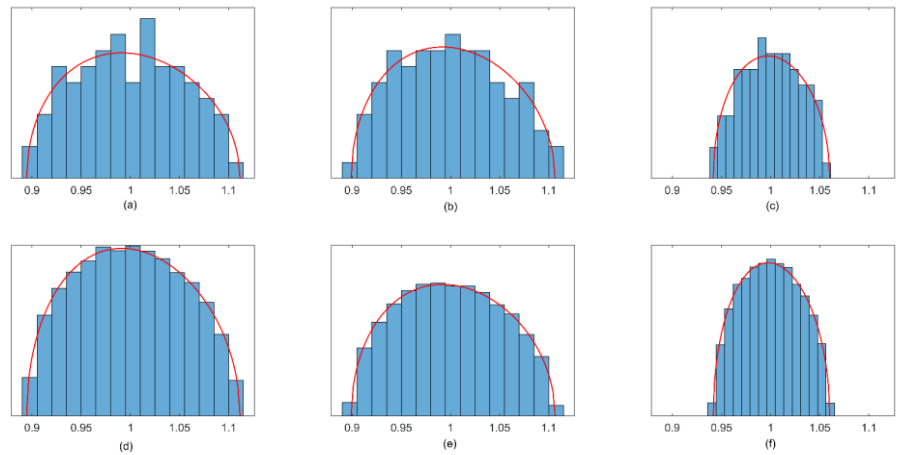

Figure 5: Bars represents eigenvector components for each stock corresponding to three largest eigenvevues for the three timespan. (a), (d) and (g) corresponds to the largest eigenvector, (b), (e ) and (h) corresponds to second largest eigenvector and (c), (f) and (i) corresponds to the third eigenvector for pre-election, election and post-election period respectively. Stocks on the x-axis are arranged according to sectors, A:autobobile, B:Consumer Goods, C:Pharmasuticals, D:Financial Services, E:Energy and F:IT sector.

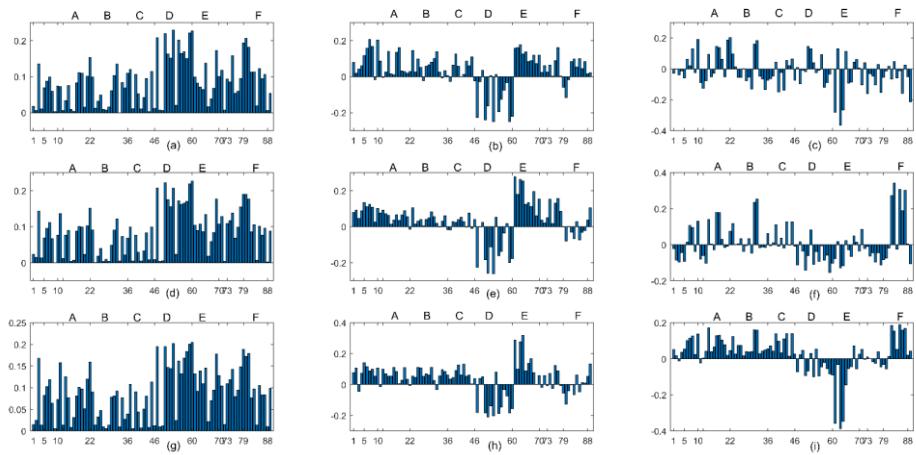

Figure 6: Network of 89 stocks based on correlation coefficient method for the period of Jan, Feb 2014 i.e. pre-election period in India. Different colours represent different sectors, also size of a node is proportional to the degree of the node and width of the edge is inversely proportional to the distance between two nodes.

Figure 7: Network of 89 stocks based on correlation coefficient method for the period of Mar, Apr, May 2014 i.e. election period in India. Different colours represent different sectors, also size of a node is proportional to the degree of the node and width of the edge is inversely proportional to the distance between two nodes

Figure 8: Network of 89 stocks based on correlation coefficient method for the period of June to Dec 2014 i.e. post-election period in India. Different colours represent different sectors, also size of a node is proportional to the degree of the node and width of the edge is inversely proportional to the distance between two nodes.

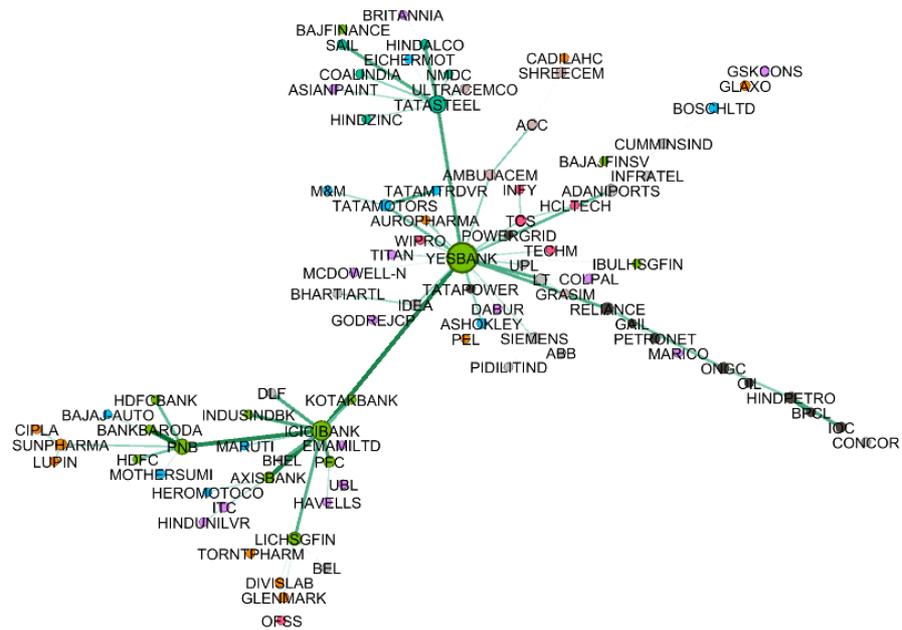

Figure 9: Network of 89 stocks based on mutual information method for the period of Jan, Feb 2014 i.e. pre-election period in India. Different colours represent different sectors, also size of a node is proportional to the degree of the node and width of the edge is inversely proportional to the distance between two nodes.

Figure 10: Network of 89 stocks based on mutual information method for the period of Mar, Apr, May 2014 i.e. election period in India. Different colours represent different sectors, also size of a node is proportional to the degree of the node and width of the edge is inversely proportional to the distance between two nodes.

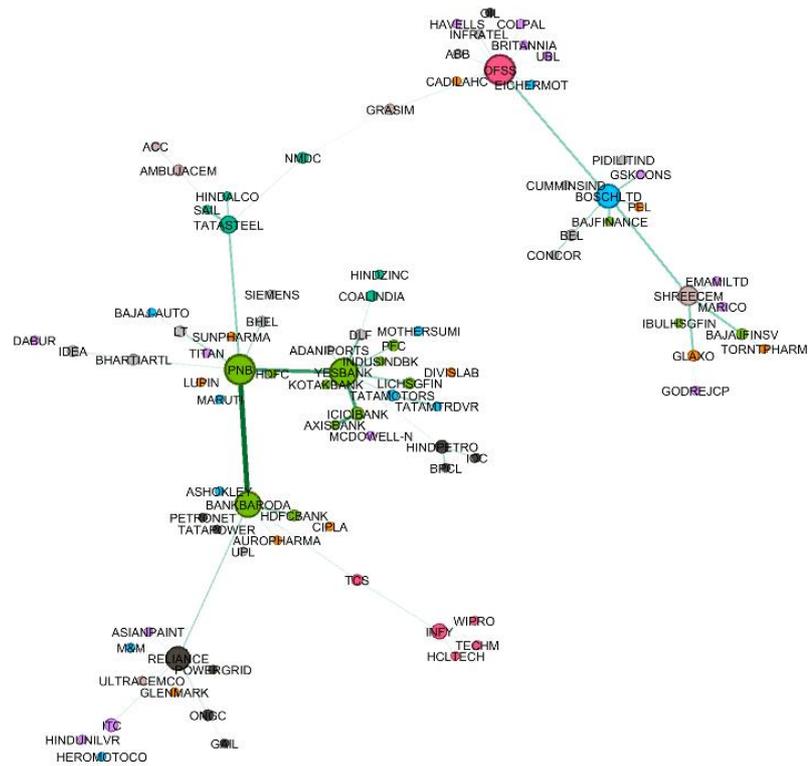

Figure 11: Network of 89 stocks based on mutual information method for the period of Jun to Dec 2014 i.e. post-election period in India. Different colours represent different sectors, also size of a node is proportional to the degree of the node and width of the edge is inversely proportional to the distance between two nodes.

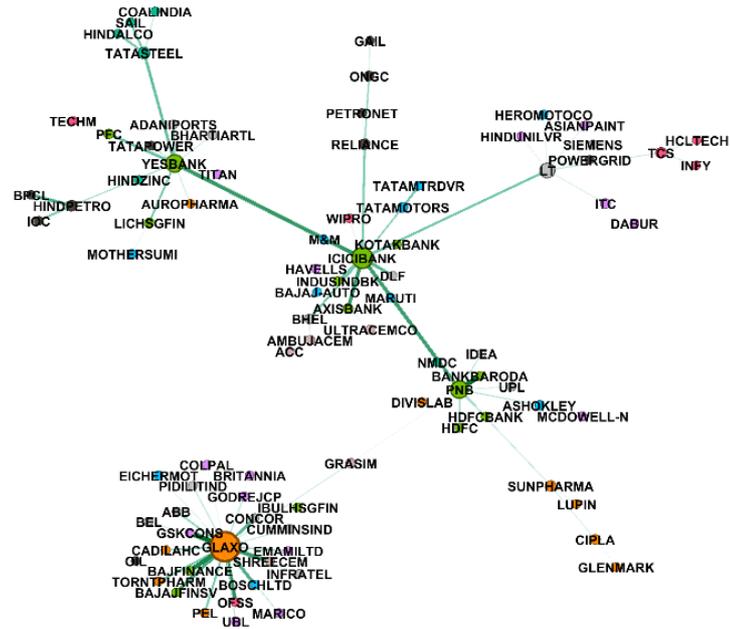

Figure 12: Normalized mutual information between all 3916 pairs versus corresponding correlation coefficient three different time span, (a) pre-election, (b) election, (c) post-election.

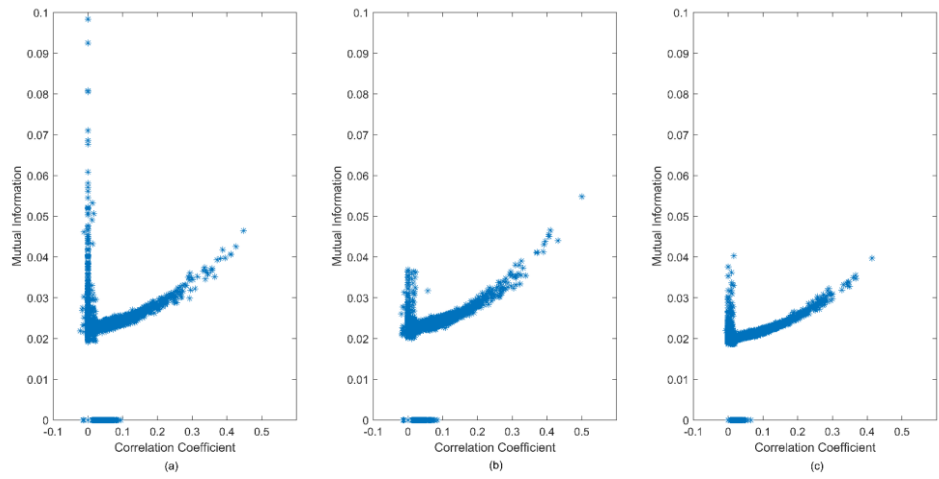

Figure 13: Degree distribution corresponding to all 12 networks.

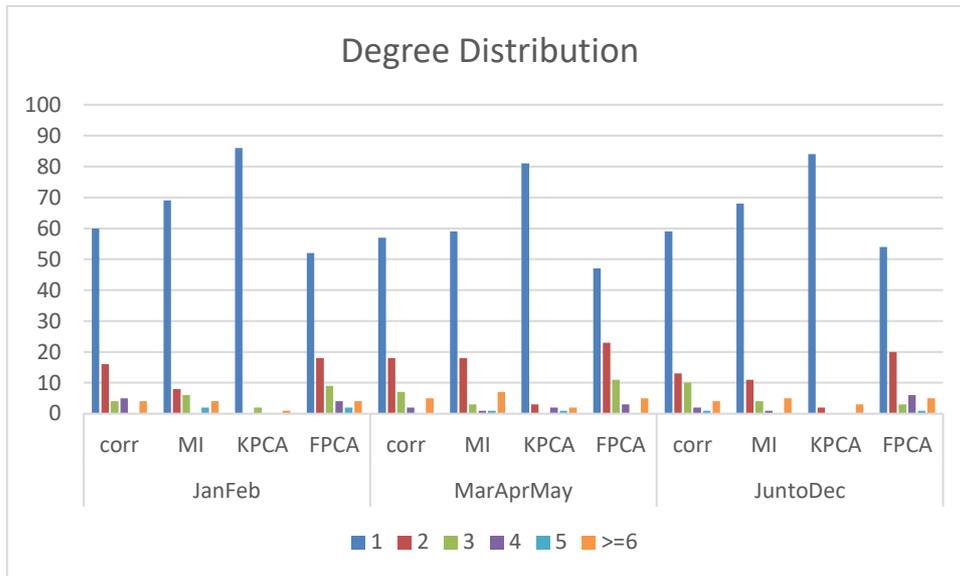